\documentclass{article}
\usepackage{amsmath}
\usepackage{amssymb}
\usepackage{array}
\usepackage{geometry}
\usepackage{graphicx}
\usepackage{enumerate}
\numberwithin{equation}{section}
\begin{document}
\begin{center}
{\bf Computable Solutions of Fractional Reaction-Diffusion Equations Associated with}\\
{\bf Generalized Riemann-Liouville Fractional Derivatives of Fractional Orders}\\

\vskip.2cm
R.K. Saxena$^1$\\
\vskip.2cm
Department of Mathematics and Statistics$^1$\\
Jai Narain Vyas University, Jodhpur-342005, India\\
ram.saxena@yahoo.com\\

\vskip.2cm
A.M. Mathai$^{2,3}$\\
\vskip.2cm
Centre for Mathematical and Statistical Sciences$^2$\\
Peechi Campus, KFRI, Peechi-680653, Kerala, India\\
directorcms458@gmail.com\\
and\\
Department of Mathematics and Statistics$^3$\\
McGill University, Montreal, Canada, H3A 2K6\\
mathai@math.mcgill.ca\\

\vskip.2cm
and\\

\vskip.2cm
H.J. Haubold$^{2,4}$\\
\vskip.2cm
Office for Outer Space Affairs, United Nations$^4$\\
Vienna International Centre, P.O. Box 500, A-1400 Vienna, Austria\\
hans.haubold@gmail.com
\end{center}

\vskip.5cm\noindent{\bf Abstract}\\
\vskip.3cm This paper is in continuation of the authors' recently published paper (Journal of Mathematical Physics 55(2014)083519) in which computational solutions of an unified reaction-diffusion equation of distributed order associated with Caputo derivatives as the time-derivative and Riesz-Feller derivative as space derivative is derived. In the present paper, computable solutions of distributed order fractional reaction-diffusion equations associated with generalized Riemann-Liouville derivatives of fractional orders as the time-derivative and Riesz-Feller fractional derivative as the space derivative are investigated. The solutions of the fractional reaction-diffusion equations of fractional orders are obtained in this paper. The method followed in deriving the solutions is that of joint Laplace and Fourier transforms. The solutions obtained are in a closed and computable form in terms of the familiar generalized Mittag-Leffler functions. They provide elegant extensions of the results given in the literature.


{\bf 1.\hskip.3cm Introduction}\\
\vskip.3cm
Distributed order fractional reaction-diffusion systems are studied, among others, by Saxton [54,55,56], Haken [16], Langlands [28], Sokolov et al. [57-59], Saxena and Pagnini [51], Guo and Xu [15], Huang and Liu [24], and recent monographs on the subject [4,22,26,27,31,38]. General models for reaction-diffusion systems are discussed by Wilhelmsson and Lazzaro [64], Henry and Wearne [19,20], Henry et al. [21], Haubold et al. [17,18], Mainardi et al. [29,30], Saxena [42], and Saxena et al. [43,44,45,47]. Stability in reaction-diffusion systems and nonlinear oscillation phenomena have been discussed by Gafiychuk et al. [12,13]. Pattern formation in reaction-diffusion related to physical and biological sciences can be found in the work of Murray [34], Cross and Hohenberg [3], and Nicolis and Prigogine [36]. Recently, Engler [7] discussed the speed of spread for fractional reaction-diffusion. Distributed order sub-diffusion is discussed by Naber [35]. General models for reaction-diffusion systems are discussed by Hilfer [22], Wilhelmsson and Lazzaro [64], Henry and Wearne [19,20], Henry et al. [21], Mainardi et al. [29,30], Haubold et al. [17,18], Diethelm [4], Kuramoto [27], Hundsdorfer and Verwer [25], and Saxena et al. [42-48]. The fundamental and numerical solution of a reaction-diffusion equation associated with the Riesz fractional derivative as the space derivative is derived by Chen et al. [2]. Reaction-diffusion models associated with Riemann-Liouville fractional derivative as the time derivative and Riesz-Feller derivative as the space derivative are recently discussed by Haubold et al. [18]. Such equations in case of Caputo fractional derivative are recently solved by Saxena et al. [50].
\vskip.2cm
In connection with the evolution equation for the probabilistic generalization of Voigt profile function, it is shown by Pagnini and Mainardi [37] that the solution of the following integro-differential equation

$$\frac{{\rm d}N}{{\rm d}t}={_0D_0}^{\alpha_1}N(x,t)+{_0D_0}^{\alpha_2}N(x,t), N(x,0)=\delta(x),\eqno(1.1)
$$is obtained in terms of the Fourier transform, where ${_0D_0}^{\alpha_1}$ and ${_0D_0}^{\alpha_2}$ are the Riesz fractional derivatives of orders $\alpha_1$ and $\alpha_2$ respectively, and $\delta(x)$ is the Dirac-delta function, which is given in [37, p.1593]. Consider the Fourier transform with parameter $k$:

$$N^{*}(k,\tau)=\exp[-\tau(|k|^{\alpha_1}+|k|^{\alpha_2})],  \tau>0.\eqno(1.2)
$$This has motivated the authors to investigate the solutions of partial differential equations (2.1), (4.1) and (5.1), listed later on. The solutions are obtained in a closed and computable form in terms of the familiar Mittag-Leffler functions. Some known and unknown results associated with fractional reaction-diffusion equations  and fractional reaction-diffusion of fractional orders can also be derived, as special cases of our findings.  It may be observed that in case of distributed order fractional reaction-diffusion equation, the solution can be written in a compact and closed form in terms of a generalization of Kamp\'e de F\'eriet hypergeometric series in two variables. Due to general character of the derived results, the results given earlier by Chen et al. [2], Haubold et al. [17] and Pagnini and Mainardi [37], Saxena et al. [48, 50], and others, readily follow as special cases of our investigations. The solutions are obtained in forms suitable for numerical computations.

\vskip.3cm\noindent{\bf 2.\hskip.3cm Solution of the Fractional Reaction-Diffusion Equation}

\vskip.3cm 
In this section, we will derive a computable solution of the one-dimensional fractional reaction-diffusion equation, given below in (2.1), containing generalized Riemann-Liouville fractional derivatives as the time-derivatives and a Riesz-Feller fractional derivative as the space derivative. The results obtained are in a compact and computable form in terms of the generalized Mittag-Leffler function, defined by (A3) in the form of the following theorem.

\vskip.3cm\noindent{\bf Theorem 1.}\hskip.3cm{\it Consider the one-dimensional fractional reaction-diffusion equation of fractional order

$$D_t^{\gamma_1,\delta_1}N(x,t)+a D_t^{\gamma_2,\delta_2}N(x,t)=\eta {_xD_{\theta}}^{\alpha}N(x,t)-\omega N(x,t)+U(x,t),\eqno(2.1)
$$Here $\omega,\eta,t>0,x\in R; \alpha,\theta,\gamma_1,\gamma_2,\delta_1,\delta_2$ are real parameters with the constraints
$$1<\gamma_1\le 2, 0\le\delta_1\le 1; 1<\gamma_2\le 2, 0\le \delta_2\le 1, 0<\alpha\le 2;\eqno(2.2)
$$ $D_t^{\gamma_1,\delta_1}$ and $D_t^{\gamma_2,\delta_2}$ are the generalized Riemann-Liouville fractional derivative operators defined by (A9) with the  conditions
\begin{align*}
I_{0_{+}}^{(1-\delta_1)(2-\gamma_1)}N(x,0)&=f_1(x); \frac{{\rm d}}{{\rm d}x}I_{0_{+}}^{(1-\delta_1)(2-\gamma_1)}N(x,0_{+})=g_1(x)\\
I_{0_{+}}^{(1-\delta_2)(2-\gamma_2)}N(x,0)&=f_2(x); \frac{{\rm d}}{{\rm d}x}I_{0_{+}}^{(1-\delta_2)(2-\gamma_2)}N(x,0)=g_2(x),\\
\lim_{|x|\to\infty}N(x,t)&=0.&(2.3)\end{align*}
Further, $\omega$ is a constant with reaction term, ${_xD_{\theta}}^{\alpha}$ is the Riesz-Feller fractional derivative of order $\alpha$ and symmetry $\theta$ defined by (A11) with $|\theta|<\min (\alpha, 2-\alpha), \eta$ is a diffusion constant and $U(x,t)$ is a nonlinear function belonging to the area of reaction-diffusion. Then the solution of (2.1), under the above conditions, is given by
\begin{align*}
N(x,t)&=t^{\gamma_1+\delta_1(2-\gamma_1)-2}\sum_{r=0}^{\infty}\frac{(-a)^r}{2\pi}\int_{-\infty}^{\infty}t^{(\gamma_1-\gamma_2)r}f_1^{*}(k)\exp(-ikx)\\
&\times E_{\gamma_1,\gamma_1+(\gamma_1-\gamma_2)r+\delta_1(2-\gamma_1)-1}^{r+1}(-bt^{\gamma_1}){\rm d}k\\
&+t^{\gamma_1+\delta_1(2-\gamma_1)-1}\sum_{r=0}^{\infty}\frac{(-a)^r}{2\pi}\int_{-\infty}^{\infty}t^{(\gamma_1-\gamma_2)r}\\
&\times g_1^{*}(k)\exp(-ikx)E_{\gamma_1,\gamma_1+(\gamma_1-\gamma_2)r+\delta_1(2-\gamma_1)}^{r+1}(-bt^{\gamma_1}){\rm d}k\\
&+at^{\gamma_1+\delta_2(2-\gamma_2)-2}\sum_{r=0}^{\infty}\frac{(-a)^r}{2\pi}\int_{-\infty}^{\infty}t^{(\gamma_1-\gamma_2)r}f_2^{*}(k)\exp(-ikx)\\
&\times E_{\gamma_1,\gamma_1+(\gamma_1-\gamma_2)r+\delta_2(2-\gamma_2)-1}^{r+1}(-bt^{\gamma_1}){\rm d}k\\
&+at^{\gamma_1+\delta_2(2-\gamma_2)-1}\sum_{r=0}^{\infty}\frac{(-a)^r}{2\pi}\int_{-\infty}^{\infty}t^{(\gamma_1-\gamma_2)r}g_2^{*}(k)\\
&\times \exp(-ikx)E_{\gamma_1,\gamma_1+(\gamma_1-\gamma_2)r+\delta_2(2-\gamma_2)}^{r+1}(-bt^{\gamma_1}){\rm d}k\\
&+\sum_{r=0}^{\infty}\frac{(-a)^r}{2\pi}\int_0^t\xi^{\gamma_1+(\gamma_1-\gamma_2)r-1}\int_{-\infty}^{\infty}U^{*}(k,t-\xi)\exp(-ikx)\\
&\times E_{\gamma_1,\gamma_1+(\gamma_1-\gamma_2)r}^{r+1}(-b\xi^{\gamma_1}){\rm d}k{\rm d}\xi,&(2.4)\end{align*}
where $\Re(\gamma_1)>0,\Re(\gamma_2)>0$ and $\Re(\gamma_1-\gamma_2)>0, b=\omega+\eta \psi_{\alpha}^{\theta}(k)$.}

\vskip.3cm\noindent{\bf Proof:}\hskip.3cm If we apply the Laplace transform with respect to the time variable and the Fourier transform with respect to the space variable $x$, use the initial conditions and the formula (A11) and (A12), then the given equation transforms into the form
\begin{align*}
s^{\gamma_1}\tilde{N}^{*}(k,s)&-s^{1-\delta_1(2-\gamma_1)}f_1^{*}(k)-s^{\delta_1(\gamma_1-2)}g_1^{*}(k)\\
&+as^{\gamma_2}\tilde{N}^{*}(k,s)-as^{1-\delta_2(2-\gamma_2)}f_2^{*}(k)-as^{\delta_2(\gamma_2-2)}g_2^{*}(k)\\
&=\eta \psi_{\alpha}^{\theta}(k)\tilde{N}^{*}(k,s)-\omega\tilde{N}^{*}(k,s)+\tilde{U}^{*}(k,s)\end{align*}
where according to the convention followed, the symbol $\backsim$ will stand for the Laplace transform with respect to the time variable $t$ and * represents the Fourier transform with respect to the space variable $x$. Solving for $\tilde{N}^{*}(k,s)$ it gives
\begin{align*}
\tilde{N}^{*}(k,s)&=\frac{1}{[s^{\gamma_1}+as^{\gamma_2}+b]}[s^{1-\delta_1(2-\gamma_1)}f_1^{*}(k)+s^{\delta_1(\gamma_1-2)}g_1^{*}(k)]\\
&+as^{1-\delta_2(2-\gamma_2)}f_2^{*}(k)+as^{\delta_2(\gamma_2-2)}g_2^{*}(k)+\tilde{U}^{*}(s)&(2.5)\end{align*}
where $b=\omega+\eta\psi_{\alpha}^{\theta}(k)$. Inverting the Laplace transform by using the result (Appendix A(19)) we obtain
\begin{align*}
\tilde{N}^{*}(k,t)&=f_1^{*}(k)t^{\gamma_1+\delta_1(2-\gamma_1)-2}\sum_{r=0}^{\infty}(-a)^rt^{(\gamma_1-\gamma_2)r}\\
&\times E_{\gamma_1,\gamma_1+(\gamma_1-\gamma_2)r+\delta_1(2-\gamma_1)-1}^{r+1}(-bt^{\gamma_1})\\
&+g_1^{*}(k)t^{\gamma_1+\delta_1(2-\gamma_1)-1}\sum_{r=0}^{\infty}(-a)^rt^{(\gamma_1-\gamma_2)r}\\
&\times E_{\gamma_1,\gamma_1+(\gamma_1-\gamma_2)r+\delta_1(2-\gamma_1)}^{r+1}(-bt^{\gamma_1})\\
&+af_2^{*}(k)t^{\gamma_1+\delta_2(2-\gamma_2)-2}\sum_{r=0}^{\infty}(-a)^rt^{(\gamma_1-\gamma_2)r}\\
&\times E_{\gamma_1,\gamma_1+(\gamma_1-\gamma_2)r+\delta_2(2-\gamma_2)-1}^{r+1}(-bt^{\gamma_1})\\
&+ag_2^{*}(k)t^{\gamma_1+\delta_2(2-\gamma_2)-1}\sum_{r=0}^{\infty}(-a)^rt^{(\gamma_1-\gamma_2)r}\\
&\times E_{\gamma_1,\gamma_1+(\gamma_1-\gamma_2)r+\delta_2(2-\gamma_2)}^{r+1}(-bt^{\gamma_1})\\
&+\int_0^tU^{*}(k,t-\xi)\sum_{r=0}^{\infty}(-a)^r\xi^{\gamma_1+(\gamma_1-\gamma_2)r-1}\\
&\times E_{\gamma_1,\gamma_1+(\gamma_1-\gamma_2)r}^{r+1}(-b\xi^{\gamma_1}).&(2.6)\end{align*}
The required solution to (2.1) is now obtained by taking the inverse Fourier transform of (2.6).

\vskip.3cm\noindent{\bf 3.\hskip.3cm Special Cases}
\vskip.3cm When $\omega =0$ Theorem 1 reduces to the following:
{\bf Corollary 1.1.}\hskip.3cm {\it Consider the one-dimensional fractional reaction-diffusion equation of fractional order
$$D_t^{\gamma_1,\delta_1}N(x,t)+a D_t^{\gamma_2,\delta_2}N(x,t)=\eta {_xD_{\theta}}^{\alpha}N(x,t)+U(x,t),\eqno(3.1)
$$for $\eta,t>0,x\in R; \alpha,\theta,\gamma_1,\gamma_2,\delta_1,\delta_2$ are real parameters with the constraints
$$1<\gamma_1\le 2, 0\le\delta_1\le 1; 1<\gamma_2\le 2, 0\le \delta_2\le 1, 0<\alpha\le 2;\eqno(3.2)
$$  with the conditions
\begin{align*}
I_{0_{+}}^{(1-\delta_1)(2-\gamma_1)}N(x,0)&=f_1(x); \frac{{\rm d}}{{\rm d}x}I_{0_{+}}^{(1-\delta_1)(2-\gamma_1)}N(x,0_{+})=g_1(x)\\
I_{0_{+}}^{(1-\delta_2)(2-\gamma_2)}N(x,0)&=f_2(x); \frac{{\rm d}}{{\rm d}x}I_{0_{+}}^{(1-\delta_2)(2-\gamma_2)}N(x,0)=g_2(x),\\
\lim_{|x|\to\infty}N(x,t)&=0.&(3.3)\end{align*}
${_xD_{\theta}}^{\alpha}$ is the Riesz-Feller fractional derivative of order $\alpha$ and symmetry $\theta$ defined by (A11) with $|\theta|<\min (\alpha, 2-\alpha), \eta$ is a diffusion constant and $U(x,t)$ is a nonlinear function belonging to the area of reaction-diffusion. Then the solution of (3.1) is given by
\begin{align*}
N(x,t)&=t^{\gamma_1+\delta_1(2-\gamma_1)-2}\sum_{r=0}^{\infty}\frac{(-a)^r}{2\pi}\int_{-\infty}^{\infty}t^{(\gamma_1-\gamma_2)r}f_1^{*}(k)\exp(-ikx)\\
&\times E_{\gamma_1,\gamma_1+(\gamma_1-\gamma_2)r+\delta_1(2-\gamma_1)-1}^{r+1}(-\eta t^{\gamma_1}\psi_{\alpha}^{\theta}(k)){\rm d}k\\
&+t^{\gamma_1+\delta_1(2-\gamma_1)-1}\sum_{r=0}^{\infty}\frac{(-a)^r}{2\pi}\int_{-\infty}^{\infty}t^{(\gamma_1-\gamma_2)r}\\
&\times g_1^{*}(k)\exp(-ikx)E_{\gamma_1,\gamma_1+(\gamma_1-\gamma_2)r+\delta_1(2-\gamma_1)}^{r+1}(-\eta t^{\gamma_1}\psi_{\alpha}^{\theta}(k)){\rm d}k\\
&+at^{\gamma_1+\delta_2(2-\gamma_2)-2}\sum_{r=0}^{\infty}\frac{(-a)^r}{2\pi}\int_{-\infty}^{\infty}t^{(\gamma_1-\gamma_2)r}f_2^{*}(k)\exp(-ikx)\\
&\times E_{\gamma_1,\gamma_1+(\gamma_1-\gamma_2)r+\delta_2(2-\gamma_2)-1}^{r+1}(-\eta t^{\gamma_1}\psi_{\alpha}^{\theta}(k)){\rm d}k\\
&+at^{\gamma_1+\delta_2(2-\gamma_2)-1}\sum_{r=0}^{\infty}\frac{(-a)^r}{2\pi}\int_{-\infty}^{\infty}t^{(\gamma_1-\gamma_2)r}g_2^{*}(k)\\
&\times \exp(-ikx)E_{\gamma_1,\gamma_1+(\gamma_1-\gamma_2)r+\delta_2(2-\gamma_2)}^{r+1}(-\eta t^{\gamma_1}\psi_{\alpha}^{\theta}(k)){\rm d}k\\
&+\sum_{r=0}^{\infty}\frac{(-a)^r}{2\pi}\int_0^t\xi^{\gamma_1+(\gamma_1-\gamma_2)r-1}\int_{-\infty}^{\infty}U^{*}(k,t-\xi)\exp(-ikx)\\
&\times E_{\gamma_1,\gamma_1+(\gamma_1-\gamma_2)r}^{r+1}(-\eta \xi^{\gamma_1}\psi_{\alpha}^{\theta}(k)){\rm d}k{\rm d}\xi,&(3.4)\end{align*}
where $\Re(\gamma_1)>0,\Re(\gamma_2)>0$ and $\Re(\gamma-\gamma_2)>0, b=\eta \psi_{\alpha}^{\theta}(k)$.}

\vskip.3cm If we set $\theta=0$ then by virtue of the relation (A14), the Riesz-Feller space derivative reduces to a Riesz derivative and Theorem 1 reduces to the following:

\vskip.3cm\noindent{\bf Corollary 1.2.}\hskip.3cm{\it Consider the one-dimensional reaction-diffusion equation of fractional order
$$D_t^{\gamma_1,\delta_1}N(x,t)+a D_t^{\gamma_2,\delta_2}N(x,t)=\eta {_xD_0}^{\alpha}N(x,t)-\omega N(x,t)+U(x,t),\eqno(3.5)
$$for $\omega,\eta,t>0,x\in R; \alpha,\theta,\gamma_1,\gamma_2,\delta_1,\delta_2$ are real parameters with the constraints
$$1<\gamma_1\le 2, 0\le\delta_1\le 1; 1<\gamma_2\le 2, 0\le \delta_2\le 1, 0<\alpha\le 2;\eqno(3.6)
$$with the conditions
\begin{align*}
I_{0_{+}}^{(1-\delta_1)(2-\gamma_1)}N(x,0)&=f_1(x); \frac{{\rm d}}{{\rm d}x}I_{0_{+}}^{(1-\delta_1)(2-\gamma_1)}N(x,0_{+})=g_1(x)\\
I_{0_{+}}^{(1-\delta_2)(2-\gamma_2)}N(x,0)&=f_2(x); \frac{{\rm d}}{{\rm d}x}I_{0_{+}}^{(1-\delta_2)(2-\gamma_2)}N(x,0)=g_2(x),\\
\lim_{|x|\to\infty}N(x,t)&=0.&(3.7)\end{align*}
$\omega$ is a constant with the reaction term, ${_xD_{0}}^{\alpha}$ is the Riesz fractional derivative of order $\alpha$ defined by (A11), $\eta$ is a diffusion constant and $U(x,t)$ is a nonlinear function belonging to the area of reaction-diffusion. Then the solution of (3.5) is the following:
\begin{align*}
N(x,t)&=t^{\gamma_1+\delta_1(2-\gamma_1)-2}\sum_{r=0}^{\infty}\frac{(-a)^r}{2\pi}\int_{-\infty}^{\infty}t^{(\gamma_1-\gamma_2)r}f_1^{*}(k)\exp(-ikx)\\
&\times E_{\gamma_1,\gamma_1+(\gamma_1-\gamma_2)r+\delta_1(2-\gamma_1)-1}^{r+1}(-\eta t^{\gamma_1}(\omega+|k|^{\alpha})){\rm d}k\\
&+t^{\gamma_1+\delta_1(2-\gamma_1)-1}\sum_{r=0}^{\infty}\frac{(-a)^r}{2\pi}\int_{-\infty}^{\infty}t^{(\gamma_1-\gamma_2)r}\\
&\times g_1^{*}(k)\exp(-ikx)E_{\gamma_1,\gamma_1+(\gamma_1-\gamma_2)r+\delta_1(2-\gamma_1)}^{r+1}(-\eta t^{\gamma_1}(\omega+|k|^{\alpha})){\rm d}k\\
&+at^{\gamma_1+\delta_2(2-\gamma_2)-2}\sum_{r=0}^{\infty}\frac{(-a)^r}{2\pi}\int_{-\infty}^{\infty}t^{(\gamma_1-\gamma_2)r}f_2^{*}(k)\exp(-ikx)\\
&\times E_{\gamma_1,\gamma_1+(\gamma_1-\gamma_2)r+\delta_2(2-\gamma_2)-1}^{r+1}(-\eta t^{\gamma_1}(\omega+|k|^{\alpha})){\rm d}k\\
&+at^{\gamma_1+\delta_2(2-\gamma_2)-1}\sum_{r=0}^{\infty}\frac{(-a)^r}{2\pi}\int_{-\infty}^{\infty}t^{(\gamma_1-\gamma_2)r}g_2^{*}(k)\\
&\times \exp(-ikx)E_{\gamma_1,\gamma_1+(\gamma_1-\gamma_2)r+\delta_2(2-\gamma_2)}^{r+1}(-\eta t^{\gamma_1}(\omega+|k|^{\alpha})){\rm d}k\\
&+\sum_{r=0}^{\infty}\frac{(-a)^r}{2\pi}\int_0^t\xi^{\gamma_1+(\gamma_1-\gamma_2)r-1}\int_{-\infty}^{\infty}U^{*}(k,t-\xi)\exp(-ikx)\\
&\times E_{\gamma_1,\gamma_1+(\gamma_1-\gamma_2)r}^{r+1}(-\eta \xi^{\gamma_1}(\omega+|k|^{\alpha})){\rm d}k{\rm d}\xi,&(3.8)\end{align*}
where $\Re(\gamma_1)>0,\Re(\gamma_2)>0$ and $\Re(\gamma-\gamma_2)>0$.}

\vskip.3cm If we set $\delta_1=\delta_2=0$ then the generalized Riemann-Liouville fractional derivatives $D_t^{\gamma_1,\delta_1}$ and $D_t^{\gamma_2,\delta_2}$ reduce respectively to the Riemann-Liouville fractional derivatives ${^{RL}_0D_t}^{\gamma_1}$ and ${^{RL}_0D_t}^{\gamma_2}$ defined by (A5) and we arrive at the following:

\vskip.3cm\noindent{\bf Corollary 1.3.}\hskip.3cm {\it Consider the one-dimensional fractional reaction-diffusion equation of fractional order
$${^{RL}_0D_t}^{\gamma_1}N(x,t)+a {^{RL}_0D_t}^{\gamma_2}N(x,t)=\eta {_xD_{\theta}}^{\alpha}N(x,t)-\omega N(x,t)+U(x,t),\eqno(3.9)
$$for $\omega,\eta,t>0,x\in R; \alpha,\theta,\gamma_1,\gamma_2,$ are real parameters with the constraints
$$1<\gamma_1\le 2; 1<\gamma_2\le 2;\eqno(3.10)
$$where ${^{RL}_0D_t}^{\gamma_1}$ and ${^{RL}_0D_t}^{\gamma_2}$ are the Riemann-Liouville fractional derivative operators defined by (A5),  with the conditions
\begin{align*}
D_t^{(\gamma_1-2)}N(x,0))&=f_1(x); D_t^{(\gamma_1-1)}N(x,0_{+})=g_1(x)\\
D_t^{(\gamma_2-2)}N(x,0)&=f_2(x); D_t^{(\gamma_2-1)}N(x,0)=g_2(x),\\
\lim_{|x|\to\infty}N(x,t)&=0.&(3.11)\end{align*}
where $D_t^{(\gamma_1-2)}N(x,0)$ denotes the $(\gamma_1-2)$th derivative of $N(x,t)$ evaluated at $t=0$ and other quantities are as defined before and the conditions on the parameters, including the one on $\theta$ remain the same. Then the solution of (3.9), under the above conditions, is given by
\begin{align*}
N(x,t)&=t^{\gamma_1-2}\sum_{r=0}^{\infty}\frac{(-a)^r}{2\pi}\int_{-\infty}^{\infty}t^{(\gamma_1-\gamma_2)r}f_1^{*}(k)\exp(-ikx)\\
&\times E_{\gamma_1,\gamma_1+(\gamma_1-\gamma_2)r-1}^{r+1}(-b t^{\gamma_1}){\rm d}k\\
&+t^{\gamma_1-1}\sum_{r=0}^{\infty}\frac{(-a)^r}{2\pi}\int_{-\infty}^{\infty}t^{(\gamma_1-\gamma_2)r}\\
&\times g_1^{*}(k)\exp(-ikx)E_{\gamma_1,\gamma_1+(\gamma_1-\gamma_2)r}^{r+1}(-b t^{\gamma_1}){\rm d}k\\
&+at^{\gamma_1-2}\sum_{r=0}^{\infty}\frac{(-a)^r}{2\pi}\int_{-\infty}^{\infty}t^{(\gamma_1-\gamma_2)r}f_2^{*}(k)\exp(-ikx)\\
&\times E_{\gamma_1,\gamma_1+(\gamma_1-\gamma_2)r-1}^{r+1}(-b t^{\gamma_1}){\rm d}k\\
&+at^{\gamma_1-1}\sum_{r=0}^{\infty}\frac{(-a)^r}{2\pi}\int_{-\infty}^{\infty}t^{(\gamma_1-\gamma_2)r}g_2^{*}(k)\\
&\times \exp(-ikx)E_{\gamma_1,\gamma_1+(\gamma_1-\gamma_2)r}^{r+1}(-b t^{\gamma_1}){\rm d}k\\
&+\sum_{r=0}^{\infty}\frac{(-a)^r}{2\pi}\int_0^t\xi^{\gamma_1+(\gamma_1-\gamma_2)r-1}\int_{-\infty}^{\infty}U^{*}(k,t-\xi)\exp(-ikx)\\
&\times E_{\gamma_1,\gamma_1+(\gamma_1-\gamma_2)r}^{r+1}(-b \xi^{\gamma_1}){\rm d}k{\rm d}\xi,&(3.12)\end{align*}
where $\Re(\gamma_1)>0,\Re(\gamma_2)>0$ and $\Re(\gamma_1-\gamma_2)>0, b=\omega+\eta \psi_{\alpha}^{\theta}(k)$.}

\vskip.3cm If we set $\delta_1=\delta_2=1$ then the generalized Riemann-Liouville fractional derivatives $D_t^{\gamma_1,\delta_1}$ and $D_t^{\gamma_2,\delta_2}$ reduced respectively to the Caputo derivatives $^C_0D_t^{\gamma_1}$ and $^C_0D_t^{\gamma_2}$ defined by (A6) and we arrive at the following:

\vskip.3cm\noindent{\bf Corollary 1.4.}\hskip.3cm {\it Consider the one-dimensional reaction-diffusion equation of fractional order
$${^C_0D_t}^{\gamma_1}N(x,t)+a {^C_0D_t}^{\gamma_2}N(x,t)=\eta {_xD_{\theta}}^{\alpha}N(x,t)-\omega N(x,t)+U(x,t)\eqno(3.13)
$$where $\omega,\eta,t>o,x\in R;\alpha,\gamma_1,\gamma_2$ are real parameters with the constraints $1<\gamma_1\le 2,1<\gamma_2\le 2,0<\alpha\le 2$ with
$$N(x,0_{+})=f(x), \frac{{\rm d}}{{\rm d}x}N(x,0_{+})=g(x))\eqno(3.14)
$$and $\omega$ is a constant with reaction term, ${_xD_{\theta}}^{\alpha}$ is the Riesz-Feller space fractional derivative of order $\alpha$ and symmetry $\theta$ defined by (A11), with $|\theta|<\min(\alpha,2-\alpha),\eta $ is the diffusion constant and $U(x,t)$ is a nonlinear function belonging to the area of reaction-diffusion. The solution of (3.13), under the above conditions, is given by the following:
\begin{align*}
N(x,t)&=\sum_{r=0}^{\infty}\frac{(-a)^r}{2\pi}\int_{-\infty}^{\infty}t^{(\gamma_1-\gamma_2)r}f_1^{*}(k)\exp(-ikx)\\
&\times E_{\gamma_1,(\gamma_1-\gamma_2)r+1}^{r+1}(-b t^{\gamma_1}){\rm d}k\\
&+\sum_{r=0}^{\infty}\frac{(-a)^r}{2\pi}\int_{-\infty}^{\infty}t^{(\gamma_1-\gamma_2)r+1}\\
&\times g_1^{*}(k)\exp(-ikx)E_{\gamma_1,(\gamma_1-\gamma_2)r+2}^{r+1}(-b t^{\gamma_1}){\rm d}k\\
&+a\sum_{r=0}^{\infty}\frac{(-a)^r}{2\pi}\int_{-\infty}^{\infty}t^{(\gamma_1-\gamma_2)(r+1)}f_2^{*}(k)\exp(-ikx)\\
&\times E_{\gamma_1,(\gamma_1-\gamma_2)(r+1)+1}^{r+1}(-b t^{\gamma_1}){\rm d}k\\
&+a\sum_{r=0}^{\infty}\frac{(-a)^r}{2\pi}\int_{-\infty}^{\infty}t^{(\gamma_1-\gamma_2)(r+1)+1}g_2^{*}(k)\\
&\times \exp(-ikx)E_{\gamma_1,(\gamma_1-\gamma_2)(r+1)+2}^{r+1}(-b t^{\gamma_1}){\rm d}k\\
&+\sum_{r=0}^{\infty}\frac{(-a)^r}{2\pi}\int_0^t\xi^{\gamma_1+(\gamma_1-\gamma_2)r-1}\int_{-\infty}^{\infty}U^{*}(k,t-\xi)\exp(-ikx)\\
&\times E_{\gamma_1,\gamma_1+(\gamma_1-\gamma_2)r}^{r+1}(-b \xi^{\gamma_1}){\rm d}k{\rm d}\xi,&(3.15)\end{align*}
where $\Re(\gamma_1)>0,\Re(\gamma_2)>0$ and $\Re(\gamma_1-\gamma_2)>0, b=\omega+\eta \psi_{\alpha}^{\theta}(k)$.}

\vskip.3cm\noindent{\bf 4.\hskip.3cm A Particularly Interesting Case}

\vskip.3cm\noindent{\bf Theorem 2.}\hskip.3cm{\it Under the conditions of Theorem 1 with $1<\gamma_1\le 2$ replaced by $1<\gamma_1<1$ and $1<\gamma_2\le 2$ replaced by $1<\gamma_2<1$ and following similar method, the solution of the following one-dimensional fractional reaction-diffusion equation of fractional order
$$D_t^{\gamma_1,\delta_1}N(x,t)+a D_t^{\gamma_2,\delta_2}N(x,t)=\eta {_xD_{\theta}}^{\alpha}N(x,t)-\omega N(x,t)+U(x,t),\eqno(4.1)
$$where $\omega,\eta,t>0,x\in R; \alpha,\theta,\gamma_1,\gamma_2,\delta_1,\delta_2$ are real parameters with the constraints
$$1<\gamma_1\le 1, 0\le\delta_1\le 1; 1<\gamma_2\le 2=1, 0\le \delta_2\le 1, 0<\alpha\le 2;\eqno(4.2)
$$with the  conditions
$$
I_{0_{+}}^{(1-\delta_1)(1-\gamma_1)}N(x,0)=h_1(x);
I_{0_{+}}^{(1-\delta_2)(1-\gamma_2)}N(x,0)=h_2(x);
\lim_{|x|\to\infty}N(x,t)=0.\eqno(4.3)$$is given by

\begin{align*}
N(x,t)&=t^{\gamma_1+\delta_1(1-\gamma_1)-1}\sum_{r=0}^{\infty}\frac{(-a)^r}{2\pi}\int_{-\infty}^{\infty}t^{(\gamma_1-\gamma_2)r}h_1^{*}(k)\exp(-ikx)\\
&\times E_{\gamma_1,\gamma_1+\delta_1+(\gamma_1-\gamma_2)r-\gamma_1\delta_1}^{r+1}(-bt^{\gamma_1}){\rm d}k\\
&+at^{\gamma_1+\delta_2(1-\gamma_2)-1}\sum_{r=0}^{\infty}\frac{(-a)^r}{2\pi}\int_{-\infty}^{\infty}t^{(\gamma_1-\gamma_2)r}h_2^{*}(k)\exp(-ikx)\\
&\times E_{\gamma_1,\gamma_1+(\gamma_1-\gamma_2)r+\delta_2(1-\gamma_2)}^{r+1}(-bt^{\gamma_1}){\rm d}k\\
&+\sum_{r=0}^{\infty}\frac{(-a)^r}{2\pi}\int_0^t\xi^{\gamma_1+(\gamma_1-\gamma_2)r-1}\int_{-\infty}^{\infty}U^{*}(k,t-\xi)\exp(-ikx)\\
&\times E_{\gamma_1,\gamma_1+(\gamma_1-\gamma_2)r}^{r+1}(-b\xi^{\gamma_1}){\rm d}k{\rm d}\xi,&(4.4)\end{align*}
where $\Re(\gamma_1)>0,\Re(\gamma_2)>0$ and $\Re(\gamma_1-\gamma_2)>0, b=\omega+\eta \psi_{\alpha}^{\theta}(k)$.}

\vskip.3cm If we set $\theta =0$, then by virtue of the relation (A14), the Riesz-Feller space derivative reduces to the Riesz derivative and Theorem 2 reduces to the following:

\vskip.3cm\noindent{\bf Corollary 2.1.}\hskip.3cm{\it Under the conditions of Theorem 2 with $\theta=0$, the solution of the following fractional one-dimensional reaction-diffusion equation of fractional order
$$D_t^{\gamma_1,\delta_1}N(x,t)+a D_t^{\gamma_2,\delta_2}N(x,t)=\eta {_xD_{0}}^{\alpha}N(x,t)-\omega N(x,t)+U(x,t),\eqno(4.4)
$$with the  conditions
$$
I_{0_{+}}^{(1-\delta_1)(1-\gamma_1)}N(x,0)=h_1(x);
I_{0_{+}}^{(1-\delta_2)(1-\gamma_2)}N(x,0)=h_2(x);
\lim_{|x|\to\infty}N(x,t)=0.\eqno(4.5)
$$is given by

\begin{align*}
N(x,t)&=t^{\gamma_1+\delta_1(1-\gamma_1)-1}\sum_{r=0}^{\infty}\frac{(-a)^r}{2\pi}\int_{-\infty}^{\infty}t^{(\gamma_1-\gamma_2)r}h_1^{*}(k)\exp(-ikx)\\
&\times E_{\gamma_1,\gamma_1+\delta_1+(\gamma_1-\gamma_2)r-\gamma_1\delta_1}^{r+1}(-bt^{\gamma_1}){\rm d}k\\
&+at^{\gamma_1+\delta_2(1-\gamma_2)-1}\sum_{r=0}^{\infty}\frac{(-a)^r}{2\pi}\int_{-\infty}^{\infty}t^{(\gamma_1-\gamma_2)r}h_2^{*}(k)\exp(-ikx)\\
&\times E_{\gamma_1,\gamma_1+(\gamma_1-\gamma_2)r+\delta_2(1-\gamma_2)}^{r+1}(-bt^{\gamma_1}){\rm d}k\\
&+\sum_{r=0}^{\infty}\frac{(-a)^r}{2\pi}\int_0^t\xi^{\gamma_1+(\gamma_1-\gamma_2)r-1}\int_{-\infty}^{\infty}U^{*}(k,t-\xi)\exp(-ikx)\\
&\times E_{\gamma_1,\gamma_1+(\gamma_1-\gamma_2)r}^{r+1}(-b\xi^{\gamma_1}){\rm d}k{\rm d}\xi,&(4.6)\end{align*}
where $\Re(\gamma_1)>0,\Re(\gamma_2)>0$ and $\Re(\gamma_1-\gamma_2)>0, b=\omega+\eta \psi_{\alpha}^{\theta}(k)$.}

\vskip.3cm\noindent{\bf Corollary 2.2.}\hskip.3cm{\it Under he conditions of Theorem 2 with $\omega=0$, the solution of the following one-dimensional fractional reaction-diffusion equation of fractional order
$$D_t^{\gamma_1,\delta_1}N(x,t)+a D_t^{\gamma_2,\delta_2}N(x,t)=\eta {_xD_{\theta}}^{\alpha}N(x,t)+U(x,t),\eqno(4.7)
$$with the  conditions
$$
I_{0_{+}}^{(1-\delta_1)(1-\gamma_1)}N(x,0)=h_1(x);
I_{0_{+}}^{(1-\delta_2)(1-\gamma_2)}N(x,0)=h_2(x);
\lim_{|x|\to\infty}N(x,t)=0.\eqno(4.8)$$is given by

\begin{align*}
N(x,t)&=t^{\gamma_1+\delta_1(1-\gamma_1)-1}\sum_{r=0}^{\infty}\frac{(-a)^r}{2\pi}\int_{-\infty}^{\infty}t^{(\gamma_1-\gamma_2)r}h_1^{*}(k)\exp(-ikx)\\
&\times E_{\gamma_1,\gamma_1+\delta_1+(\gamma_1-\gamma_2)r-\gamma_1\delta_1}^{r+1}(-b^{*}t^{\gamma_1}){\rm d}k\\
&+at^{\gamma_1+\delta_2(1-\gamma_2)-1}\sum_{r=0}^{\infty}\frac{(-a)^r}{2\pi}\int_{-\infty}^{\infty}t^{(\gamma_1-\gamma_2)r}h_2^{*}(k)\exp(-ikx)\\
&\times E_{\gamma_1,\gamma_1+(\gamma_1-\gamma_2)r+\delta_2(1-\gamma_2)}^{r+1}(-b^{*}t^{\gamma_1}){\rm d}k\\
&+\sum_{r=0}^{\infty}\frac{(-a)^r}{2\pi}\int_0^t\xi^{\gamma_1+(\gamma_1-\gamma_2)r-1}\int_{-\infty}^{\infty}U^{*}(k,t-\xi)\exp(-ikx)\\
&\times E_{\gamma_1,\gamma_1+(\gamma_1-\gamma_2)r}^{r+1}(-b^{*}\xi^{\gamma_1}){\rm d}k{\rm d}\xi,&(4.9)\end{align*}
where $\Re(\gamma_1)>0,\Re(\gamma_2)>0$ and $\Re(\gamma_1-\gamma_2)>0, b^{*}=\eta \psi_{\alpha}^{\theta}(k)$.}

\vskip.3cm If we set $\delta_1=0,\delta_2=0$ then the generalized Riemann-Liouville fractional derivatives $D_t^{\gamma_1,\delta_1}$ and $D_t^{\gamma_2,\delta_2}$ reduce to Riemann-Liouville fractional derivatives ${^{RL}_0D_t}^{\gamma_1}$ and ${^{RL}_0D_t}^{\gamma_2}$ defined in (A5) and we arrive at the following:

\vskip.3cm\noindent{\bf Corollary 2.3.}\hskip.3cm{\it Under the conditions of Theorem 2 with $\delta_1=0=\delta_2$, the solution of one-dimensional reaction-diffusion equation of fractional order
$${^{RL}_0D_t}^{\gamma_1}N(x,t)+a {^{RL}_0D_t}^{\gamma_2}N(x,t)=\eta {_xD_{\theta}}^{\alpha}N(x,t)-\omega N(x,t)+U(x,t),\eqno(4.10)
$$for $\omega,\eta,t>0,x\in R; \alpha,\theta,\gamma_1,\gamma_2,$ are real parameters with the constraints
with the conditions
\begin{align*}
D_t^{(\gamma_1-1)}N(x,0))&=e_1(x); D_t^{(\gamma_1-2)}N(x,0_{+})=e_2(x)\\
\lim_{|x|\to\infty}N(x,t)&=0, t>0.&(4.11)\end{align*}
is given by
\begin{align*}
N(x,t)&=t^{\gamma_1-1}\sum_{r=0}^{\infty}\frac{(-a)^r}{2\pi}\int_{-\infty}^{\infty}t^{(\gamma_1-\gamma_2)r}e_1^{*}(k)\exp(-ikx)\\
&\times E_{\gamma_1,\gamma_1+(\gamma_1-\gamma_2)r}^{r+1}(-b t^{\gamma_1}){\rm d}k\\
&+at^{\gamma_1-1}\sum_{r=0}^{\infty}\frac{(-a)^r}{2\pi}\int_{-\infty}^{\infty}t^{(\gamma_1-\gamma_2)r}\\
&\times e_2^{*}(k)\exp(-ikx)E_{\gamma_1,\gamma_1+(\gamma_1-\gamma_2)r}^{r+1}(-b t^{\gamma_1}){\rm d}k\\
&+\sum_{r=0}^{\infty}\frac{(-a)^r}{2\pi}\int_0^t\xi^{\gamma_1+(\gamma_1-\gamma_2)r-1}\int_{-\infty}^{\infty}U^{*}(k,t-\xi)\exp(-ikx)\\
&\times E_{\gamma_1,\gamma_1+(\gamma_1-\gamma_2)r}^{r+1}(-b \xi^{\gamma_1}){\rm d}k{\rm d}\xi,&(4.12)\end{align*}
where $\Re(\gamma_1)>0,\Re(\gamma_2)>0$ and $\Re(\gamma_1-\gamma_2)>0, b=\omega+\eta \psi_{\alpha}^{\theta}(k)$.}

\vskip.3cm If $\delta_1=\delta_2=1$, the generalized Riemann-Liouville fractional derivatives $D_t^{\gamma_1,\delta_1}$  and $D_t^{\gamma_2,\delta_2}$ reduce respectively to the Caputo fractional derivatives ${^C_0D_t}^{\gamma_1}$ and ${^C_0D_t}^{\gamma_2}$ defined by (A6) and we arrive at the following:

\vskip.3cm\noindent{\bf Corollary 2.4.}\hskip.3cm{\it Under he conditions of Theorem 2 with $h_1(x)=h_2(x)=h(x)$ and $\delta_1=\delta_2=1$, the solution of the following one-dimensional fractional reaction-diffusion equation of fractional order
$${^C_0D_t}^{\gamma_1}N(x,t)+a {^C_0D_t}^{\gamma_2}N(x,t)=\eta {_xD_{\theta}}^{\alpha}N(x,t)-\omega N(x,t)+U(x,t)\eqno(4.13)
$$with the constraints
$$N(x,0_{+})=h(x), \lim_{|x|\to\infty}N(x,t)=0.\eqno(4.14)
$$is given by the following:
\begin{align*}
N(x,t)&=\sum_{r=0}^{\infty}\frac{(-a)^r}{2\pi}\int_{-\infty}^{\infty}t^{(\gamma_1-\gamma_2)r}h^{*}(k)\exp(-ikx)\\
&\times E_{\gamma_1,(\gamma_1-\gamma_2)r+1}^{r+1}(-b t^{\gamma_1}){\rm d}k\\
&+a\sum_{r=0}^{\infty}\frac{(-a)^r}{2\pi}\int_{-\infty}^{\infty}t^{(\gamma_1-\gamma_2)(r+1)}\\
&\times h^{*}(k)\exp(-ikx)E_{\gamma_1,(\gamma_1-\gamma_2)(r+1)+1}^{r+1}(-b t^{\gamma_1}){\rm d}k\\
&+\sum_{r=0}^{\infty}\frac{(-a)^r}{2\pi}\int_0^t\xi^{\gamma_1+(\gamma_1-\gamma_2)r-1}\int_{-\infty}^{\infty}U^{*}(k,t-\xi)\exp(-ikx)\\
&\times E_{\gamma_1,\gamma_1+(\gamma_1-\gamma_2)r}^{r+1}(-b \xi^{\gamma_1}){\rm d}k{\rm d}\xi,&(4.15)\end{align*}
where $\Re(\gamma_1)>0,\Re(\gamma_2)>0$ and $\Re(\gamma_1-\gamma_2)>0, b=\omega+\eta \psi_{\alpha}^{\theta}(k)$.}

\vskip.3cm For $\omega=0$ (4.13) reduces to one given by Saxena et al. [50]. If we further set $h(x)=\delta(x)$, where $\delta(x)$ is the Dirac-delta function, then Theorem 2 can be written in the following form:

\vskip.3cm\noindent{\bf Corollary 2.5.}\hskip.3cm{\it Under the conditions of Theorem 2, the fundamental solution of the following one-dimensional reaction-diffusion equation of fractional order
$${^C_0D_t}^{\gamma_1}N(x,t)+a {^C_0D_t}^{\gamma_2}N(x,t)=\eta {_xD_{\theta}}^{\alpha}N(x,t)-\omega N(x,t)+U(x,t)\eqno(4.16)
$$with the constraints
$$N(x,0_{+})=\delta(x), \lim_{|x|\to\infty}N(x,t)=0, t>0.\eqno(4.17)
$$is given by the following:
\begin{align*}
N(x,t)&=\sum_{r=0}^{\infty}\frac{(-a)^r}{2\pi}\int_{-\infty}^{\infty}t^{(\gamma_1-\gamma_2)r}\exp(-ikx)\\
&\times E_{\gamma_1,(\gamma_1-\gamma_2)r+1}^{r+1}(-b t^{\gamma_1}){\rm d}k\\
&+a\sum_{r=0}^{\infty}\frac{(-a)^r}{2\pi}\int_{-\infty}^{\infty}t^{(\gamma_1-\gamma_2)(r+1)}\\
&\times \exp(-ikx)E_{\gamma_1,(\gamma_1-\gamma_2)(r+1)+1}^{r+1}(-b t^{\gamma_1}){\rm d}k\\
&+\sum_{r=0}^{\infty}\frac{(-a)^r}{2\pi}\int_0^t\xi^{\gamma_1+(\gamma_1-\gamma_2)r-1}\int_{-\infty}^{\infty}U^{*}(k,t-\xi)\exp(-ikx)\\
&\times E_{\gamma_1,\gamma_1+(\gamma_1-\gamma_2)r}^{r+1}(-b \xi^{\gamma_1}){\rm d}k{\rm d}\xi,&(4.18)\end{align*}
where $\Re(\gamma_1)>0,\Re(\gamma_2)>0$ and $\Re(\gamma_1-\gamma_2)>0, b=\omega+\eta \psi_{\alpha}^{\theta}(k)$.}

\vskip.3cm For $\omega=0$, (4.17) yields another result given by Saxena et al. [50]. As a concluding remark, it is interesting to observe that Theorem 2 also hold true if instead of one Riesz-Feller derivative, we consider a finite number of Feller derivatives. This result is given in Theorem 3 in the next section.\vskip.3cm\noindent{\bf 5.\hskip.3cm Several Riesz-Feller Space Fractional Derivatives}

\vskip.3cm\noindent{\bf Theorem 3.}\hskip.3cm{\it Consider the one-dimensional fractional reaction-diffusion equation of fractional order
$$D_t^{\gamma_1,\delta_1}N(x,t)+a D_t^{\gamma_2,\delta_2}N(x,t)=\sum_{j=1}^m[\eta_j {_xD_{\theta_j}}^{\alpha_j}N(x,t)]-\omega N(x,t)+U(x,t),\eqno(5.1)
$$where $\omega,\eta_j,t>0,x\in R; \alpha_j,\theta_j,j=1,...,m,\gamma_1,\gamma_2,\delta_1,\delta_2$ are real parameters with the constraints
$$1<\gamma_1\le 2, 0\le\delta_1\le 1; 1<\gamma_2\le 2, 0\le \delta_2\le 1, 0<\alpha_j\le 2, j=1,...,m;\eqno(5.2)
$$ $D_t^{\gamma_1,\delta_1}$ and $D_t^{\gamma_2,\delta_2}$ are the generalized Riemann-Liouville fractional derivative operators defined by (A9) with the  conditions
\begin{align*}
I_{0_{+}}^{(1-\delta_1)(2-\gamma_1)}N(x,0)&=f_1(x); \frac{{\rm d}}{{\rm d}x}I_{0_{+}}^{(1-\delta_1)(2-\gamma_1)}N(x,0_{+})=g_1(x)\\
I_{0_{+}}^{(1-\delta_2)(2-\gamma_2)}N(x,0)&=f_2(x); \frac{{\rm d}}{{\rm d}x}I_{0_{+}}^{(1-\delta_2)(2-\gamma_2)}N(x,0)=g_2(x),\\
\lim_{|x|\to\infty}N(x,t)&=0.&(5.3)\end{align*}
Further, $\omega$ is a constant with reaction term, ${_xD_{\theta_j}}^{\alpha_j}$ are the Riesz-Feller fractional derivatives of order $\alpha_j$ and symmetry $\theta_j$ defined by (A11) with $|\theta_j|<\min (\alpha_j, 2-\alpha_j), j=1,...,m, \eta_j>0,j=1,...,m$ are the diffusion constants and $U(x,t)$ is a nonlinear function belonging to the area of reaction-diffusion. Then the solution of (5.1), under the above conditions, is given by
\begin{align*}
N(x,t)&=t^{\gamma_1+\delta_1(2-\gamma_1)-2}\sum_{r=0}^{\infty}\frac{(-a)^r}{2\pi}\int_{-\infty}^{\infty}t^{(\gamma_1-\gamma_2)r}f_1^{*}(k)\exp(-ikx)\\
&\times E_{\gamma_1,\gamma_1+(\gamma_1-\gamma_2)r+\delta_1(2-\gamma_1)-1}^{r+1}(-\hat{b}t^{\gamma_1}){\rm d}k\\
&+t^{\gamma_1+\delta_1(2-\gamma_1)-1}\sum_{r=0}^{\infty}\frac{(-a)^r}{2\pi}\int_{-\infty}^{\infty}t^{(\gamma_1-\gamma_2)r}\\
&\times g_1^{*}(k)\exp(-ikx)E_{\gamma_1,\gamma_1+(\gamma_1-\gamma_2)r+\delta_1(2-\gamma_1)}^{r+1}(-\hat{b}t^{\gamma_1}){\rm d}k\\
&+at^{\gamma_1+\delta_2(2-\gamma_2)-2}\sum_{r=0}^{\infty}\frac{(-a)^r}{2\pi}\int_{-\infty}^{\infty}t^{(\gamma_1-\gamma_2)r}f_2^{*}(k)\exp(-ikx)\\
&\times E_{\gamma_1,\gamma_1+(\gamma_1-\gamma_2)r+\delta_2(2-\gamma_2)-1}^{r+1}(-\hat{b}t^{\gamma_1}){\rm d}k\\
&+at^{\gamma_1+\delta_2(2-\gamma_2)-1}\sum_{r=0}^{\infty}\frac{(-a)^r}{2\pi}\int_{-\infty}^{\infty}t^{(\gamma_1-\gamma_2)r}g_2^{*}(k)\\
&\times \exp(-ikx)E_{\gamma_1,\gamma_1+(\gamma_1-\gamma_2)r+\delta_2(2-\gamma_2)}^{r+1}(-\hat{b}t^{\gamma_1}){\rm d}k\\
&+\sum_{r=0}^{\infty}\frac{(-a)^r}{2\pi}\int_0^t\xi^{\gamma_1+(\gamma_1-\gamma_2)r-1}\int_{-\infty}^{\infty}U^{*}(k,t-\xi)\exp(-ikx)\\
&\times E_{\gamma_1,\gamma_1+(\gamma_1-\gamma_2)r}^{r+1}(-\hat{b}\xi^{\gamma_1}){\rm d}k{\rm d}\xi,&(5.4)\end{align*}
where $\Re(\gamma_1)>0,\Re(\gamma_2)>0$ and $\Re(\gamma_1-\gamma_2)>0, \hat{b}=\omega+\sum_{j=1}^m\eta_j \psi_{\alpha_j}^{\theta_j}(k)$.}

\vskip.3cm\noindent{\bf 6.\hskip.3cm Special Cases of Theorem 3}

\vskip.3cm\noindent{\bf Corollary 3.1.}\hskip.3cm{\it Under the conditions of Theorem 3, with $\omega=0$, the one-dimensional fractional reaction-diffusion equation of fractional order
$$D_t^{\gamma_1,\delta_1}N(x,t)+a D_t^{\gamma_2,\delta_2}N(x,t)=\sum_{j=1}^m[\eta_j {_xD_{\theta_j}}^{\alpha_j}N(x,t)]+U(x,t),\eqno(6.1)
$$where $\eta_j,t>0,x\in R; \alpha_j,\theta_j,j=1,...,m,\gamma_1,\gamma_2,\delta_1,\delta_2$ are real parameters with the constraints
$$1<\gamma_1\le 2, 0\le\delta_1\le 1; 1<\gamma_2\le 2, 0\le \delta_2\le 1, 0<\alpha_j\le 2, j=1,...,m;\eqno(6.2)
$$ $D_t^{\gamma_1,\delta_1}$ and $D_t^{\gamma_2,\delta_2}$ are the generalized Riemann-Liouville fractional derivative operators defined by (A9) with the  conditions
\begin{align*}
I_{0_{+}}^{(1-\delta_1)(2-\gamma_1)}N(x,0)&=f_1(x); \frac{{\rm d}}{{\rm d}x}I_{0_{+}}^{(1-\delta_1)(2-\gamma_1)}N(x,0_{+})=g_1(x)\\
I_{0_{+}}^{(1-\delta_2)(2-\gamma_2)}N(x,0)&=f_2(x); \frac{{\rm d}}{{\rm d}x}I_{0_{+}}^{(1-\delta_2)(2-\gamma_2)}N(x,0)=g_2(x),\\
\lim_{|x|\to\infty}N(x,t)&=0.&(6.3)\end{align*}
Further, ${_xD_{\theta_j}}^{\alpha_j}$ are the Riesz-Feller fractional derivatives of order $\alpha_j$ and symmetry $\theta_j$ $|\theta_j|<\min (\alpha_j, 2-\alpha_j), j=1,...,m, \eta_j>0,j=1,...,m$ are the diffusion constants and $U(x,t)$ is a nonlinear function belonging to the area of reaction-diffusion. Then the solution of (6.1), under the above conditions, is given by
\begin{align*}
N(x,t)&=t^{\gamma_1+\delta_1(2-\gamma_1)-2}\sum_{r=0}^{\infty}\frac{(-a)^r}{2\pi}\int_{-\infty}^{\infty}t^{(\gamma_1-\gamma_2)r}f_1^{*}(k)\exp(-ikx)\\
&\times E_{\gamma_1,\gamma_1+(\gamma_1-\gamma_2)r+\delta_1(2-\gamma_1)-1}^{r+1}(-\tilde{b}t^{\gamma_1}){\rm d}k\\
&+t^{\gamma_1+\delta_1(2-\gamma_1)-1}\sum_{r=0}^{\infty}\frac{(-a)^r}{2\pi}\int_{-\infty}^{\infty}t^{(\gamma_1-\gamma_2)r}\\
&\times g_1^{*}(k)\exp(-ikx)E_{\gamma_1,\gamma_1+(\gamma_1-\gamma_2)r+\delta_1(2-\gamma_1)}^{r+1}(-\tilde{b}t^{\gamma_1}){\rm d}k\\
&+at^{\gamma_1+\delta_2(2-\gamma_2)-2}\sum_{r=0}^{\infty}\frac{(-a)^r}{2\pi}\int_{-\infty}^{\infty}t^{(\gamma_1-\gamma_2)r}f_2^{*}(k)\exp(-ikx)\\
&\times E_{\gamma_1,\gamma_1+(\gamma_1-\gamma_2)r+\delta_2(2-\gamma_2)-1}^{r+1}(-\tilde{b}t^{\gamma_1}){\rm d}k\\
&+at^{\gamma_1+\delta_2(2-\gamma_2)-1}\sum_{r=0}^{\infty}\frac{(-a)^r}{2\pi}\int_{-\infty}^{\infty}t^{(\gamma_1-\gamma_2)r}g_2^{*}(k)\\
&\times \exp(-ikx)E_{\gamma_1,\gamma_1+(\gamma_1-\gamma_2)r+\delta_2(2-\gamma_2)}^{r+1}(-\tilde{b}t^{\gamma_1}){\rm d}k\\
&+\sum_{r=0}^{\infty}\frac{(-a)^r}{2\pi}\int_0^t\xi^{\gamma_1+(\gamma_1-\gamma_2)r-1}\int_{-\infty}^{\infty}U^{*}(k,t-\xi)\exp(-ikx)\\
&\times E_{\gamma_1,\gamma_1+(\gamma_1-\gamma_2)r}^{r+1}(-\tilde{b}\xi^{\gamma_1}){\rm d}k{\rm d}\xi,&(6.4)\end{align*}
where $\Re(\gamma_1)>0,\Re(\gamma_2)>0$ and $\Re(\gamma_1-\gamma_2)>0, \tilde{b}=\sum_{j=1}^m\eta_j \psi_{\alpha_j}^{\theta_j}(k)$.}

\vskip.3cm When $\theta_1=0=\theta_2=...=\theta_m$ the Riesz-Feller space derivatives reduce to Riesz Fractional derivative, then by virtue of the relation (A14), there holds the following:

\vskip.3cm\noindent{\bf Corollary 3.2.}\hskip.3cm {\it Under the conditions of Theorem 3, the fractional reaction-diffusion equation of fractional order
$$D_t^{\gamma_1,\delta_1}N(x,t)+a D_t^{\gamma_2,\delta_2}N(x,t)=\sum_{j=1}^m[\eta_j {_xD_{0}}^{\alpha_j}N(x,t)]-\omega N(x,t)+U(x,t),\eqno(6.5)
$$where $\omega, t>0,x\in R; \eta_j>0, \alpha_j, j=1,...,m,\gamma_1,\gamma_2,\delta_1,\delta_2$ are real parameters with the constraints
$$1<\gamma_1\le 2, 0\le\delta_1\le 1; 1<\gamma_2\le 2, 0\le \delta_2\le 1, 0<\alpha_j\le 2, j=1,...,m;\eqno(6.6)
$$ $D_t^{\gamma_1,\delta_1}$ and $D_t^{\gamma_2,\delta_2}$ are the generalized Riemann-Liouville fractional derivative operators defined by (A9) with the  conditions
\begin{align*}
I_{0_{+}}^{(1-\delta_1)(2-\gamma_1)}N(x,0)&=f_1(x); \frac{{\rm d}}{{\rm d}x}I_{0_{+}}^{(1-\delta_1)(2-\gamma_1)}N(x,0_{+})=g_1(x)\\
I_{0_{+}}^{(1-\delta_2)(2-\gamma_2)}N(x,0)&=f_2(x); \frac{{\rm d}}{{\rm d}x}I_{0_{+}}^{(1-\delta_2)(2-\gamma_2)}N(x,0)=g_2(x),\\
\lim_{|x|\to\infty}N(x,t)&=0.&(6.7)\end{align*}
Then the solution of (6.1), under the above conditions, is given by
\begin{align*}
N(x,t)&=t^{\gamma_1+\delta_1(2-\gamma_1)-2}\sum_{r=0}^{\infty}\frac{(-a)^r}{2\pi}\int_{-\infty}^{\infty}t^{(\gamma_1-\gamma_2)r}f_1^{*}(k)\exp(-ikx)\\
&\times E_{\gamma_1,\gamma_1+(\gamma_1-\gamma_2)r+\delta_1(2-\gamma_1)-1}^{r+1}(-b^{*}t^{\gamma_1}){\rm d}k\\
&+t^{\gamma_1+\delta_1(2-\gamma_1)-1}\sum_{r=0}^{\infty}\frac{(-a)^r}{2\pi}\int_{-\infty}^{\infty}t^{(\gamma_1-\gamma_2)r}\\
&\times g_1^{*}(k)\exp(-ikx)E_{\gamma_1,\gamma_1+(\gamma_1-\gamma_2)r+\delta_1(2-\gamma_1)}^{r+1}(-b^{*}t^{\gamma_1}){\rm d}k\\
&+at^{\gamma_1+\delta_2(2-\gamma_2)-2}\sum_{r=0}^{\infty}\frac{(-a)^r}{2\pi}\int_{-\infty}^{\infty}t^{(\gamma_1-\gamma_2)r}f_2^{*}(k)\exp(-ikx)\\
&\times E_{\gamma_1,\gamma_1+(\gamma_1-\gamma_2)r+\delta_2(2-\gamma_2)-1}^{r+1}(-b^{*}t^{\gamma_1}){\rm d}k\\
&+at^{\gamma_1+\delta_2(2-\gamma_2)-1}\sum_{r=0}^{\infty}\frac{(-a)^r}{2\pi}\int_{-\infty}^{\infty}t^{(\gamma_1-\gamma_2)r}g_2^{*}(k)\\
&\times \exp(-ikx)E_{\gamma_1,\gamma_1+(\gamma_1-\gamma_2)r+\delta_2(2-\gamma_2)}^{r+1}(-b^{*}t^{\gamma_1}){\rm d}k\\
&+\sum_{r=0}^{\infty}\frac{(-a)^r}{2\pi}\int_0^t\xi^{\gamma_1+(\gamma_1-\gamma_2)r-1}\int_{-\infty}^{\infty}U^{*}(k,t-\xi)\exp(-ikx)\\
&\times E_{\gamma_1,\gamma_1+(\gamma_1-\gamma_2)r}^{r+1}(-b^{*}\xi^{\gamma_1}){\rm d}k{\rm d}\xi,&(6.8)\end{align*}
where $\Re(\gamma_1)>0,\Re(\gamma_2)>0$ and $\Re(\gamma_1-\gamma_2)>0, b^{*}=\omega+\sum_{j=1}^m\eta_j \psi_{\alpha_j}^{\theta_j}(k)$.}

\vskip.3cm If we set $\delta_1=\delta_2=0$, then the generalized Riemann-Liouville fractional derivatives ${^{RL}_0D_t}^{\gamma_1}$ and ${^{RL}_0D_t}^{\gamma_2}$ defined by (A5) and we arrive at the following result:

\vskip.3cm\noindent{\bf Corollary 3.3.}\hskip.3cm{\it Consider the one-dimensional fractional reaction-diffusion equation of fractional order
$${^{RL}_0D_t}^{\gamma_1,\delta_1}N(x,t)+a {^{RL}_0D_t}^{\gamma_2,\delta_2}N(x,t)=\sum_{j=1}^m[\eta_j {_xD_{\theta_j}}^{\alpha_j}N(x,t)]-\omega N(x,t)+U(x,t),\eqno(6.9)
$$where $\omega, t>0,x\in R; \eta_j>0, \alpha_j, j=1,...,m,\gamma_1,\gamma_2$ are real parameters with the constraints
$$1<\gamma_1\le 2,  1<\gamma_2\le 2, 0<\alpha_j\le 2, j=1,...,m;\eqno(6.10)
$$ ${^{RL}_0D_t}^{\gamma_1,\delta_1}$ and ${^{RL}_0D_t}^{\gamma_2,\delta_2}$ are the generalized Riemann-Liouville fractional derivative operators defined by (A5) with the  conditions
\begin{align*}
D_t^{\gamma_1-2}N(x,0)&=f_1(x); D_t^{\gamma_1-1}N(x,0_{+})=g_1(x)\\
D_t^{\gamma_2-2}N(x,0)&=f_2(x); D_1^{\gamma_2-1}N(x,0)=g_2(x),\\
\lim_{|x|\to\infty}N(x,t)&=0&(6.11)\end{align*}
where $D_t^(\gamma_j-2)N(x,0)$ denotes the $(\gamma_j-2)$th derivative of $N(x,t)$ evaluated at $t=0$, $\omega$ is a constant with reaction terms, ${_xD_{\theta_j}}^{\alpha_j}$ are the Riesz-Feller space fractional derivatives f order $\alpha_j$ and symmetries $\theta_j$; $|\theta_j|<\min(\alpha_j,2-\alpha_j),j=1,...,m$ defined by (A11), $\eta_j>0$ are diffusion constants and $U(x,t)$ is a conlinear function belonging to the area of reaction-diffusion. Then the solution of (6.9), under the  above conditions, is given by
\begin{align*}
N(x,t)&=t^{\gamma_1-2}\sum_{r=0}^{\infty}\frac{(-a)^r}{2\pi}\int_{-\infty}^{\infty}t^{(\gamma_1-\gamma_2)r}f_1^{*}(k)\exp(-ikx)\\
&\times E_{\gamma_1,\gamma_1+(\gamma_1-\gamma_2)r-1}^{r+1}(-qt^{\gamma_1}){\rm d}k\\
&+t^{\gamma_1-1}\sum_{r=0}^{\infty}\frac{(-a)^r}{2\pi}\int_{-\infty}^{\infty}t^{(\gamma_1-\gamma_2)r}\\
&\times g_1^{*}(k)\exp(-ikx)E_{\gamma_1,\gamma_1+(\gamma_1-\gamma_2)r}^{r+1}(-qt^{\gamma_1}){\rm d}k\\
&+at^{\gamma_1-2}\sum_{r=0}^{\infty}\frac{(-a)^r}{2\pi}\int_{-\infty}^{\infty}t^{(\gamma_1-\gamma_2)r}f_2^{*}(k)\exp(-ikx)\\
&\times E_{\gamma_1,\gamma_1+(\gamma_1-\gamma_2)r-1}^{r+1}(-qt^{\gamma_1}){\rm d}k\\
&+at^{\gamma_1-1}\sum_{r=0}^{\infty}\frac{(-a)^r}{2\pi}\int_{-\infty}^{\infty}t^{(\gamma_1-\gamma_2)r}g_2^{*}(k)\\
&\times \exp(-ikx)E_{\gamma_1,\gamma_1+(\gamma_1-\gamma_2)r}^{r+1}(-qt^{\gamma_1}){\rm d}k\\
&+\sum_{r=0}^{\infty}\frac{(-a)^r}{2\pi}\int_0^t\xi^{\gamma_1+(\gamma_1-\gamma_2)r-1}\int_{-\infty}^{\infty}U^{*}(k,t-\xi)\exp(-ikx)\\
&\times E_{\gamma_1,\gamma_1+(\gamma_1-\gamma_2)r}^{r+1}(-q\xi^{\gamma_1}){\rm d}k{\rm d}\xi,&(6.12)\end{align*}
where $\Re(\gamma_1)>0,\Re(\gamma_2)>0$ and $\Re(\gamma_1-\gamma_2)>0, q=\sum_{j=1}^m\eta_j |k|^{\alpha_j}$.}\vskip.3cm For $\omega =0$ the above result reduces to the result given by Saxena et al. [48]. If we set $\delta_1=\delta_2=1$ then the generalized Riemann-Liouville fractional derivatives $D_t^{\gamma_1,\delta_1}$ and $D_t^{\gamma_2,\delta_2}$ reduce respectively  to the Caputo fractional derivatives ${^C_0D_t}^{\gamma_1}$ and ${^C_0D_t}^{\gamma_2}$ defined by (A6) and we arrive at the following:

\vskip.3cm\noindent{\bf Corollary 3.4.}\hskip.3cm{\it Under the conditions of Theorem 3, the solution of one-dimensional fractional reaction-diffusion equation of fractional order

$${^C_0D_t}^{\gamma_1}N(x,t)+a {^C_0D_t}^{\gamma_2}N(x,t)=[\sum_{j=1}^m\eta_j {_xD_{\theta_j}}^{\alpha_j}]N(x,t)-\omega N(x,t)+U(x,t)\eqno(6.13)
$$where $\omega,t>0; \eta_j>0,\alpha_j,\theta_j,j=1,...,m,\gamma_1,\gamma_2$ are real parameters with the conditions
$1<\gamma_1\le 2,1<\gamma_2\le 2,0<\alpha_j\le 2$ and
with the constraints
$$N(x,0_{+})=f(x);  \frac{{\rm d}}{{\rm d}x}N(x,0_{+})=g(x)\eqno(6.14)
$$and other quantities are as defined in Theorem 3. Then the solution of (6.13) is given by the following:
\begin{align*}
N(x,t)&=\sum_{r=0}^{\infty}\frac{(-a)^r}{2\pi}\int_{-\infty}^{\infty}t^{(\gamma_1-\gamma_2)r}f^{*}(k)\exp(-ikx)\\
&\times E_{\gamma_1,(\gamma_1-\gamma_2)r+1}^{r+1}(-b^{*} t^{\gamma_1}){\rm d}k\\
&+t\sum_{r=0}^{\infty}\frac{(-a)^r}{2\pi}\int_{-\infty}^{\infty}t^{(\gamma_1-\gamma_2)r}g^{*}(k)\exp(-ikx)\\
&\times E_{\gamma_1,(\gamma_1-\gamma_2)r+2}^{r+1}(-b^{*} t^{\gamma_1}){\rm d}k\\
&+a\sum_{r=0}^{\infty}\frac{(-a)^r}{2\pi}\int_{-\infty}^{\infty}t^{(\gamma_1-\gamma_2)(r+1)}f^{*}(k)\exp(-ikx)\\
&\times E_{\gamma_1,(\gamma_1-\gamma_2)(r+1)+1}^{r+1}(-b^{*} t^{\gamma_1}){\rm d}k\\
&+a\sum_{r=0}^{\infty}\frac{(-a)^r}{2\pi}\int_{-\infty}^{\infty}t^{(\gamma_1-\gamma_2)(r+1)+1}\\
&\times g^{*}(k) \exp(-ikx)E_{\gamma_1,(\gamma_1-\gamma_2)(r+1)+2}^{r+1}(-b t^{\gamma_1}){\rm d}k\\
&+\sum_{r=0}^{\infty}\frac{(-a)^r}{2\pi}\int_0^t\xi^{\gamma_1+(\gamma_1-\gamma_2)r-1}\int_{-\infty}^{\infty}U^{*}(k,t-\xi)\exp(-ikx)\\
&\times E_{\gamma_1,\gamma_1+(\gamma_1-\gamma_2)r}^{r+1}(-b^{*} \xi^{\gamma_1}){\rm d}k{\rm d}\xi,&(6.15)\end{align*}
where $\Re(\gamma_1)>0,\Re(\gamma_2)>0$ and $\Re(\gamma_1-\gamma_2)>0, b^{*}=\omega+\sum_{j=1}^m\eta_j \psi_{\alpha_j}^{\theta_j}(k)$.}

\vskip.3cm\noindent{\bf 7. \hskip.3cm Conclusions}\\

\vskip.3cm In this paper, the authors have presented the solutions of three one-dimensional fractional reaction-diffusion equations of fractional orders associated with generalized Riemann-Liouville derivative as the time derivative due to Hilfer et al. [23] and Riesz-Feller derivatives as the space derivatives. The results obtained provide an elegant extension of the solution of one-dimensional fractional reaction-diffusion equations associated with Caputo fractional derivatives as the time derivatives and Riesz-Feller fractional derivatives as the space derivatives [50]. The results are obtained in terms of generalized Mittag-Leffler functions. Further, the derived results include, as a special case, the results for the solution of space-time fractional reaction-diffusion systems associated with a generalized Riemann-Liouville fractional derivative given by the authors [49], which itself is a generalization of the fundamental solution of space-time fractional diffusion given by Mainardi et al. [29]. The solution of the equations (2.1),(4.1), and (5.1) are obtained in terms of generalized Mittag-Leffler functions. The importance of the results obtained in this paper further lies in the fact that due to presence of the modified Hilfer fractional derivative [23], results for Riemann-Liouville fractional derivative and Caputo fractional derivatives can be deduced as special cases by taking $\nu=0$ and $\delta_1=0$ and $\delta_2=1$ respectively.

\vskip.3cm\noindent{\bf Acknowledgement}

\vskip.3cm The authors would like to thank the Department of Science and Technology, Government of India, for the financial assistance for this work under project No.SR/S4/MS:287/05.

\vskip.3cm\noindent{\bf References}

\vskip.2cm\noindent [1]\hskip.3cm Caputo, M.: Elasticita e Dissipazione, Zanichelli, Bologna, 1969.

\vskip.2cm\noindent [2]\hskip.3cm Chen, J., Liu, R., Turner, I. and Anh, V.: The fundamental and numerical solutions of the Riesz space fractional reaction-dispersion equation, The Australian and New Zealand Industrial and Applied Mathematics Journal (ANZIAM) 50 (2008), 45-57.

\vskip.2cm\noindent [3]\hskip.3cm Cross, M.C. and Hohenberg, P.C.: Pattern formation outside of equilibrium, Rev. Modern Phys. 65(1993)851-912.

\vskip.2cm\noindent [4]\hskip.3cm Diethelm, K.: The Analysis of Fractional Differential Equations, Springer, Berlin, 2010.

\vskip.2cm\noindent [5]\hskip.3cm Dzherbashyan, M.M.: Integral Transforms and Representations of Functions in Complex Domain (in Russian), Nauka, Moscow, 1966.

\vskip.2cm\noindent [7]\hskip.3cm Engler, H.: On the speed of spread for fractional reaction-diffusion, International Journal of Differential Equations, Volume 2010, Article ID 315421, 16 pages.

\vskip.2cm\noindent [8]\hskip.3cm Erd\'elyi, A., Magnus, W., Oberhettinger, F. and Tricomi, F.G.: Tables of Integral Transforms, Vol. 1, McGraw-Hill, New York, 1954.
\vskip.2cm\noindent [9]\hskip.3cm Erd\'elyi, A., Magnus, W., Oberhettinger, F. and Tricomi, F.G.: Higher Transcendental Functions, Vol. 3, McGraw-Hill, New York, 1955.

\vskip.2cm\noindent [10]\hskip.3cm Feller, W.: On a generalization of Marcel Riesz' potentials and the semi-groups generated by them, Meddelenden Lunds Universitets Matematiska Seminariu(Comm. S\'em. Math\'em. Universit\'e de Lund), Tome Supple. D\'ed\'e \'a M. Riesz, Lund, 73-81(1952).

\vskip.2cm\noindent [11]\hskip.3cm Feller, W.: An Introduction to Probability Theory and Its Applications, Vol. 2, Second Edition, Wiley, New York, 1971.

\vskip.2cm\noindent [12]\hskip.3cm Gafiychuk, V., Datsko, B. and Meleshko, V.: Mathematical modeling in pattern formation in sub and super-diffusive reaction-diffusion systems, arXiv:nlin.AO/0811005v3.

\vskip.2cm\noindent [13]\hskip.3cm Gafiychuk, V., Datsko, B. and Meleshko, V.: Nonlinear oscillations and stability domains in fractional reaction-diffusion systems, arXiv:nlinPS/0702013v1.

\vskip.2cm\noindent [14]\hskip.3cm Gorenflo, R. and Mainardi, F.: Approximation of Levy-Feller diffusion by random walk, Journal for Analysis and Its Applications 18(1999), No.2, 1-16.

\vskip.2cm\noindent [15]\hskip.3cm Guo, X. and Xu, M.: Some physical applications of Schr\"odinger equation, J. Math. Phy. 47082104(2008), doi:10,1063/1.2235026 (9 pages).
\vskip.2cm\noindent [16]\hskip.3cm Haken, H.: Synergetics: Introduction and Advanced Topics, Springer, Berlin, 2004.

\vskip.2cm\noindent [17]\hskip.3cm Haubold, H.J., Mathai, A.M. and Saxena, R.K.: Solutions of the reaction-diffusion equations in terms of the H-function, Bulletin Astro. Soc., India 35(2007)681-689.

\vskip.2cm\noindent [18]\hskip.3cm Haubold, H.J., Mathai, A.M. and Saxena, R.K.: Further solutions of reaction-diffusion equations in terms of the H-function, J. Comput. Appl. Math. 235(2011)1311-1316.

\vskip.2cm\noindent [19]\hskip.3cm Henry, B.I. and Wearne, S.L.: Fractional reaction-diffusion, Physica A 276(2000)448-455.

\vskip.2cm\noindent [20]\hskip.3cm Henry, B.I. and Wearne, S.L.: Existence of Turing instabilities in a two-species fractional reaction-diffusion system, SIAM J. Appl. Math. 62(2002)870-887.

\vskip.2cm\noindent [21]\hskip.3cm Henry, B.I., Langlands, T.A.M. and Wearne, S.L.: Turing pattern formation in fractional activator-inhibitor systems, Physical Review E 72(2005)026101.

\vskip.2cm\noindent [22]\hskip.3cm Hilfer, R.: Fractional time evolution, In: Hilfer, R. (Editor), Applications of Fractional Calculus in Physics, World Scientific Publishing, Singapore, 2000, pp. 87-130.

\vskip.2cm\noindent [23]\hskip.3cm Hilfer, R., Luchko, Y. and Tomovski, Z.: Operational method for the solution of fractional differential equations with generalized Riemann-Liouville fractional derivatives, Fract. Calc. Appl. Anal. 12(2009) 299-318.

\vskip.2cm\noindent [24]\hskip.3cm Huang, F. and Liu, R.: The time-fractional diffusion equation and the advection-dispersion equation, The Australian and New Zealand Industrial and Applied Mathematics Journal (ANZIAM) 46(2005)1-14.

\vskip.2cm\noindent [25]\hskip.3cm Hundsdorfer, W. and Verwer, J.G.: Numerical Solution of Time-dependent Advection-diffusion-reaction Equations, Springer-Verlag, Berlin, 2003.

\vskip.2cm\noindent [26]\hskip.3cm Kilbas, A.A., Srivastava, H.M. and Trujillo, J.J.: Theory and Applications of Fractional Differential Equations, Elsevier, Amsterdam, 2006.

\vskip.2cm\noindent [27]\hskip.3cm Kuramoto, Y.I.: Chemical Oscillations, Waves and Turbulence, Dover Publications, Mineola, New York, 2003.

\vskip.2cm\noindent [28]\hskip.3cm Langlands, T.A.M.: Solution of a modified fractional diffusion equation, Physica A 367(2006)136-144.

\vskip.2cm\noindent [29]\hskip.3cm Mainardi, F., Luchko, Y. and Pagnini, G.: The fundamental solution of the space-time fractional diffusion equation, Fract. Calc. Appl. Math., 4(2001)153-192.

\vskip.2cm\noindent [30]\hskip.3cm Mainardi, F., Pagnini, G. and Saxena, R.K.: Fox H-function in fractional diffusion, J. COmput. Appl. Math. 178(2005)321-331.

\vskip.2cm\noindent [31]\hskip.3cm Mathai, A.M., Saxena, R.K. and Haubold, H.J.: The H-function: Theory and Applications, Springer, New York, (2010).

\vskip.2cm\noindent [32]\hskip.3cm Mittag-Leffler, G.M.: Sur la nouvelle fonction $E_{\alpha}(x)$, C.R. Acad. Sci., Paris (Ser.II) 137(1903)554-558.

\vskip.2cm\noindent [33]\hskip.3cm Mittag-Leffler, G.M.: Sur la representation analytique d'une fonction branche uniforme d'une fonction, Acta Math. 239(1905)101-181.

\vskip.2cm\noindent [34]\hskip.3cm Murray, J.D.: Mathematical Biology, Springer-Verlag, New York, 2003.

\vskip.2cm\noindent [35]\hskip.3cm Naber, M.: Distributed order fractional sub-diffusion, Fractals 12(2004)23-32.

\vskip.2cm\noindent [36]\hskip.3cm Nicolis, G. and Prigogine, I.: Self-organization in Nonequilibrium Systems: From Dissipative Structures to Order Through Fluctuations, Wiley, New York, 1977.

\vskip.2cm\noindent [37]\hskip.3cm Pagnini, R. and Mainardi, F.: Evolution equations for a probabilistic generalization of Voigt profile function, J. Comput. Appl. Math. 233(2010)1590-1595.

\vskip.2cm\noindent [38]\hskip.3cm Podlubny, I.: Fractional Differential Equations, Academic Press, New York, 1999.

\vskip.2cm\noindent [39]\hskip.3cm Prabhakar, T.R.: A singular integral equation with a generalized Mittag-Leffler function in kernel, Yokohama Math.J. 19(1971)7-15.

\vskip.2cm\noindent [40]\hskip.3cm Samko, S.G., Kilbas, A.M. and Marichev, O.I.: Fractional Integrals and Derivatives: Theory and Applications, Gordon and Breach, New York, 1990.

\vskip.2cm\noindent [41]\hskip.3cm Sandev, R., Metzler, R. and Tomovski, Z.: Fractional diffusion equation with a generalized Riemann-Liouville time fractional derivative, J. Phys. A: Math. Theor. 44(2011)255201.

\vskip.2cm\noindent [42]\hskip.3cm Saxena, R.K.: Solutions of fractional partial differential equations related to quantum mechanics, Algebra Groups and Geometries 29(2012)147-164.

\vskip.2cm\noindent [43]\hskip.3cm Saxena, R.K., Mathai, A.M. and Haubold, H.J.: Fractional reaction-diffusion equations,  Astrophysics and Space Science 305(2006a)289-296.

\vskip.2cm\noindent [44]\hskip.3cm Saxena, R.K., Mathai, A.M. and Haubold, H.J.: Reaction-diffusion systems and nonlinear waves, Astrophysics and Space Science 305(2006b)297-303.

\vskip.2cm\noindent [45]\hskip.3cm Saxena, R.K., Mathai, A.M. and Haubold, H.J.: Solutions of fractional reaction-diffusion equations in terms of the Mittag-Leffler functions Int. J. Sci. Res. 15(2006c)1-17.

\vskip.2cm\noindent [46]\hskip.3cm Saxena, R.K., Mathai, A.M. and Haubold, H.J.: Distributed order reaction-diffusion systems associated with Caputo derivatives, arXiv:1109.4841v1[math-ph].

\vskip.2cm\noindent [47]\hskip.3cm Saxena, R.K., Mathai, A.M. and Haubold, H.J.: Computational solution of unified fractional reaction-diffusion equations with composite fractional time derivative, arXiv:1210.1453v1[math-ph].

\vskip.2cm\noindent [48]\hskip.3cm Saxena, R.K., Mathai, A.M. and Haubold, H.J.: Computational solution of distributed order reaction-diffusion systems associated with Riemann-Liouville derivatives, arXiv:1211.0063v1[math-ph].

\vskip.2cm\noindent [49]\hskip.3cm Saxena, R.K., Mathai, A.M. and Haubold, H.J.: Space-time fractional reaction-diffusion equations associated with a generalized Riemann-Liouville derivative, Axioms 2014.3.320.334; doi:10.3390/axioms 3030320.

\vskip.2cm\noindent [50]\hskip.3cm Saxena, R.K., Mathai, A.M. and Haubold, H.J.: Distributed reaction-diffusion systems associated with Caputo derivative, J. Math. Phys. 55(2014)083519.

\vskip.2cm\noindent [51]\hskip.3cm Saxena, R.K. and Pagnini, G.: Exact solutions of triple order time-fractional differential equations for anomalous relaxation and diffusion: the accelerating case, Physica A 390(2011)602-613.

\vskip.2cm\noindent [52]\hskip.3cm Saxena, R.K., Saxena, R. and Kalla, S.L.: Solution of space-time fractional Schr\"odinger equation occurring in quantum mechanics, Fractional Calculus and Applied Analysis 13(2012)177-190.

\vskip.2cm\noindent [53]\hskip.3cm Saxena, R.K., Tomovsku, Z. and Sandev, T.: Fractional Helmholtz and fractional wave equations with Riesz-Feller and Riemann-Lioouville fractional derivatives, 7(2014)312-334.

\vskip.2cm\noindent [54]\hskip.3cm Saxton, M.: Anomalous diffusion due to obstacles: a Monte Carlo study, Biophys. J. 66(1994)394-401.

\vskip.2cm\noindent [55]\hskip.3cm Saxton, M.: Anomalous diffusion due to binding: a Monte Carlo study, Biophys. J. 70(1996)1250-1262.

\vskip.2cm\noindent [56]\hskip.3cm Saxton, M.: Anomalous sub-diffusion in flourescence photo bleaching recovery: a Monte Carlo study, 81(2001)2226-2240.

\vskip.2cm\noindent [57]\hskip.3cm Sokolov, I.M., Chechkin, A.V. and Klafter, J.: Distributed-order fractional kinetics, Acta Phys. Pol. B 35(2004) 1323-1341.

\vskip.2cm\noindent [58]\hskip.3cm  Sokolov, I.M. and Klater, J.: From diffusion to anomalous diffusion: a century after Einstein's Brownian motion, Chaos 15(2005)026103.

\vskip.2cm\noindent [59]\hskip.3cm Srivastava, H.M. and Tomovski, Z.: Fractional calculus with an integral operator containing a generalized Mittag-Leffler function in the kernel, Appl. Math. Comput. 21(2010)198-210.

\vskip.2cm\noindent [60]\hskip.3cm Tomovski, Z.: Generalized Cauchy type problems for nonlinear fractional differential equation with composite fractional derivative operator, Nonlinear Analysis Volume 2012, doi:10.1016/j.na.

\vskip.2cm\noindent [61]\hskip.3cm Tomovski, A., Hilfer, R. and Srivastava, H.M.: Fractional and operational calculus with generalized fractional derivative operators and Mittag-Leffler type functions, Integral Transforms and Special Functions 21(2010)797-814.

\vskip.2cm\noindent [62]\hskip.3cm Tomovski, Z., Sandev, T., Metzler, R. and Dubbeldam, J.: Generalized space-time fractional diffusion equation with composite fractional time derivative, Physica A 391(2012)2527-2542.

\vskip.2cm\noindent [63]\hskip.3cm Wilhelmsson, H. and Lazzaro, E.: Reaction-diffusion Problems in the Physics of Hot Plasmas, Institute of Physics Publishing, Bristol and Philadelphia, 2001.

\vskip.2cm\noindent [64]\hskip.3cm Wiman, A.: Ueber den Fundamentalsatz in der Theorie der Functionen $E_{\alpha}(x)$, Acta Math. 29(1905)191-201.

\vskip.3cm\noindent{\bf Appendix: Mathematical Preliminaries}

\vskip.3cm A generalization of the Mittag-Leffler function (Mittag-Leffler [32,33])

$$E_{\alpha}(z)=\sum_{n=0}^{\infty}\frac{z^n}{\Gamma(n\alpha+1)}, \alpha\in C, \Re(\alpha)>0\eqno(A1)
$$was introduced by Wiman [64] in the generalized form

$$E_{\alpha,\beta}(z)=\sum_{n=0}^{\infty}\frac{z^n}{\Gamma(n\alpha+\beta)},\Re(\alpha)>0,\Re(\beta)>0.\eqno(A2)
$$A further generalization of the Mittag-Leffler function is given by Prabhakar [39] in the form

$$E_{\alpha,\beta}^{\gamma}(z)=\sum_{n=0}^{\infty}\frac{(\gamma)_nz^n}{\Gamma(n\alpha+\beta)n!},\Re(\alpha)>0,\Re(\beta)>0,\eqno(A3)
$$where the Pochhammer symbol is given by
$$(a)_n=a(a+1)...(a+n-1), a\ne 0, (a)_0=1.
$$The main results of the Mittag-Leffler functions defined by (A1) and (A2) are available in the handbook of Erd\'elyi et al. [9, Section 18.1] and the monographs written by Dzherbashyan [5,6]. The left-sided Riemann-Liouville fractional integral of order $\nu$ is defined by Samko et al. [40], Kilbas et al. [26], Mathai et al. [31] as
$${^{RL}_0D_t}^{-\nu}N(x,t)=\frac{1}{\Gamma(\nu)}\int_0^t(t-u)^{\nu-1}N(x,u){\rm d}u, t>0, \Re(\nu)>0.\eqno(A4)
$$The left-sided Riemann-Liouville fractional derivative of order $\alpha$ is defined as

$${^{RL}_0D_t}^{\mu}N(x,t)=(\frac{{\rm d}}{{\rm d}x})^nI_0^{n-\mu}N(x,t), \Re(\mu)>0, n=[\Re(\mu)]+1,\eqno(A5)
$$where $[x]$ represents the greatest integer in the real number $x$. Caputo fractional derivative operator (Capuato[1]) is defined in the form

\begin{align*}
{^C_0D_t}^{\alpha}f(x,t)&=\frac{1}{\Gamma(m-\alpha)}\int_0^t\frac{f^{(m)}(x,\tau){\rm d}\tau}{(t-\tau)^{ \alpha+1-m}},m-1<\alpha<m\Re(\alpha)>0,m\in N&(A6)\\
&=\frac{\partial^m f(x,t)}{\partial t^m},\ mbox{ if }\alpha=m&(A7)\end{align*}
where $\frac{\partial^m}{\partial t^m}f(x,t)$ is the $m$th partial derivative of $f(x,t)$ with respect to $t$. When there is no confusion, then the Caputo operator ${C_0D_t}^{\alpha}$ will be simply denoted by ${_0D_t^{\alpha}}$. A generalization of the Riemann-Liouville fractional derivative operator (A5) as well as Caputo fractional derivative operator (A6) is given by Hilfer [22] by introducing a left-sided fractional derivative operator of two parameters of order $0<\mu< 1$ and type $0\le \nu\ 1$ in the form

$$D_{a_{+}}^{\mu,\nu}N(x,t)=\left[I_{a_{+}}^{\nu(1-\mu)}\frac{\partial}{\partial x}(I_{a_{+}}^{(1-\nu)(1-\mu)}N(x,t))\right].\eqno(A8)
$$For $\nu=0$, (A8) reduces to the classical Riemann-Liouville fractional derivative operator (A4). On the other hand, for $\nu=1$ it yields the Caputo fractional derivative operator defined by (A5).
\vskip.3cm\noindent{\bf Note A1:}\hskip.3cm The derivative defined by (A8) also occurs in recent papers by Hilfer et al. [23], Srivastava et al. [59], Tomovski [60], Tomovski [61,62] and Saxena et al. [52]. Recently, Hilfer operator defined by (A8) is rewritten in a more general form (Hilfer et al. [23]) as

\begin{align*}
D_{a_{+}}^{\mu,\nu}N(x,t)&=\left[I_{a_{+}}^{\nu(n-\nu)}\frac{{\rm d}^n}{{\rm d}x^n}(I_{a_{+}}^{(1-\nu)(n-\mu)}N(x,t))\right]\\
&=\left[I_{a_{+}}^{\nu(n-\mu)}(D_{a_{+}}^{\mu+\nu n-\mu\nu)}N(x,t)\right]&(A9)\end{align*}
Where $n-1<\mu\leq n,0\leq\nu\leq1$. The Laplace transform of the above operator (A9) is given by Tomovski [59,p.17] in the following form:
\begin{align*}
L[D_{a_{+}}^{\mu,\nu}N(x,t);s]&=s^{\mu}\tilde{N}(x,s)-\sum_{k=0}^{n-1}s^{n-k-\nu(n-\mu)-1}\\
&\times\frac{{\rm d}^k}{{\rm  d}x^k}(I_{0_{+}}^{(1-\nu)(n-\mu)}N(x,0_{+}),&(A10)
\end{align*} where $n-1<\mu\le n, n\in N,0\le\nu \le 1$. Following Feller [10,11], it is conventional to define the Riesz-Feller space fractional derivative of order $\alpha$ and skewness $\theta$ in terms of its Fourier transform as

$$F\{{_xD_{\theta}}^{\alpha}f(x);k\}=-\psi_{\alpha}^{\theta}(k)f^{*}(k),\eqno(A11)
$$where
$$\psi_{\alpha}^{\theta}(k)=|k|^{\alpha}\exp[i(sign k)\frac{\theta\pi}{2}],0<\alpha\le 2,|\theta|\le \min(\alpha,2-\alpha).\eqno(A12)
$$When $\theta=0$, we have a symmetric operator with respect to $x$, that can be interpreted as
$${_xD_0}^{\alpha}=-(-\frac{{\rm d}^2}{{\rm d}x^2})^{\frac{\alpha}{2}}.\eqno(A13)
$$This can be formally deduced by writing $-(k)^{\alpha}=-(k^2)^{\frac{\alpha}{2}}$. For $\theta=0$, we also have
$$F\{{_xD_0}^{\alpha}f(x);k\}=-|k|^{\alpha}f^{*}(k).\eqno(A14)
$$For $0<\alpha\le 2$ and $|\theta|\le \min(\alpha,2-\alpha)$, the Riesz-Feller derivative can be shown to possess the following integral representation in $x$ domain

\begin{align*}
{_xD_{\theta}}^{\alpha}f(x)&=\frac{\Gamma(1+\alpha)}{\pi}[\sin((\alpha+\theta)\pi/2)\int_0^{\infty}\frac{f(x+\xi)-f(x)}{\xi^{1+\alpha}}{\rm d}\xi\\
&+\sin((\alpha-\theta)\pi/2)\int_0^{\infty}\frac{f(x-\xi)-f(x)}{\xi^{1+\alpha}}{\rm d}\xi].\end{align*}
For $\theta=0$, the Riesz-Feller fractional derivative becomes the Riesz fractional derivative of order $\alpha$ for $1<\alpha\le 2$ defined by analytic continuation in the whole range $0<\alpha \le 2$, $\alpha\ne 1$ (see Gorenflo and Mainardi [14]) as
$${_xD_0}^{\alpha}=-\lambda[I_{+}^{-\alpha}-I_{-}^{-\alpha}]\eqno(A15)
$$where
$$\lambda=\frac{1}{2\cos(\alpha \pi/2)}; I_{\pm}^{-\alpha}=\frac{{\rm d}^2}{{\rm f}x^2}I_{\pm}^{2-\alpha}.\eqno(A16)
$$The Weyl fractional integral operators are defined in the monograph by Samko et al. [40] as
\begin{align*}
(I_{+}^{\beta}N)(x)&=\frac{1}{\Gamma(\beta)}\int_{-\infty}^x(x-\xi)^{\beta-1}N(\xi){\rm d}\xi,\beta>0\\
(I_{-}^{\beta}N)(x)&=\frac{1}{\Gamma(\beta)}\int_x^{\infty}(\xi-x)^{\beta-1}N(\xi){\rm d}\xi,\beta>o.&(A17)\end{align*}

\vskip.3cm\noindent{\bf Note A2}:\hskip.3cm We note that ${_xD_0}^{\alpha}$ is a pseudo differential operator. In particular, we have
$${_xD_0}^{\alpha}=\frac{{\rm d}^2}{{\rm d}x^2}, ~\mbox{but}~{_xD_0}^{1}\neq \frac{{\rm d}}{{\rm d}x}.\eqno(A18)
$$
The following result given by Saxena et al. [44] is also required:

$$L^{-1}[\frac{s^{\rho-1}}{s^{\alpha}+as^{\beta}+b};t]=t^{\alpha-\rho}\sum_{r=0}^{\infty}(-a)^rt^{(\alpha-\beta)r}E_{\alpha,\alpha+(\alpha-\beta)
r-\rho+1}^{r+1}(-bt^{\alpha})\eqno(A19)
$$where $\Re(\alpha)>0,\Re(\beta)>0,\Re(\alpha-\beta)>0,\Re(\alpha-\rho)>0,\Re(s)>0, |\frac{as^{\beta}}{s^{\alpha}+b}|<1$ and $E_{\alpha,\beta}^{\gamma}(z)$ is the generalized Mittag-Leffler function defined by (A3).

\end{document}